\documentclass[10pt,twocolumn,letterpaper]{article}

\usepackage{wacv}
\usepackage{times}
\usepackage{epsfig}
\usepackage{graphicx}
\usepackage{amsmath}
\usepackage{amssymb}

\def\wacvPaperID{657} % Enter the WACV Paper ID here

\wacvfinalcopy % *** Uncomment this line for the final submission

\ifwacvfinal
\usepackage[breaklinks=true,bookmarks=false]{hyperref}
\else
\usepackage[pagebackref=true,breaklinks=true,colorlinks,bookmarks=false]{hyperref}
\fi

\begin{document}

%%%%%%%%% TITLE
\title{Hyper-Convolution Networks for Biomedical Image Segmentation}

\author{Tianyu Ma\thanks{This work is accepted to WACV 2022}\\
Cornell University\\
Cornell Tech\\
{\tt\small tm478@cornell.edu}
\and
Adrian V. Dalca\\
Massachusetts Institute of Technology\\
Massachusetts General Hospital\\
Harvard Medical School\\
{\tt\small adalca@mit.edu}
\and
Mert R. Sabuncu\\
Cornell University\\
Cornell Tech\\
{\tt\small msabuncu@cornell.edu}
}
\maketitle
%\thispagestyle{empty}

%%%%%%%%% ABSTRACT
\begin{abstract}
The convolution operation is a central building block of neural network architectures widely used in computer vision.
The size of the convolution kernels determines both the expressiveness of convolutional neural networks (CNN), as well as the number of learnable parameters. 
Increasing the network capacity to capture rich pixel relationships requires increasing the number of learnable parameters, often leading to overfitting and/or lack of robustness.
In this paper, we propose a powerful novel building block, the \textit{hyper-convolution}, which implicitly represents the convolution kernel as a function of kernel coordinates. 
Hyper-convolutions enable decoupling the kernel size, and hence its receptive field, from the number of learnable parameters.
In our experiments, focused on challenging biomedical image segmentation tasks, we demonstrate that replacing regular convolutions with hyper-convolutions leads to more efficient architectures that achieve improved accuracy. 
Our analysis also shows that learned hyper-convolutions are naturally regularized, which can offer better generalization performance. 
We believe that hyper-convolutions can be a powerful building block in future neural network architectures for computer vision tasks.
We provide all of our code here: \emph{https://github.com/tym002/Hyper-Convolution}
\end{abstract}

%%%%%%%%% BODY TEXT
\section{Introduction}

Deep convolutional neural networks (CNNs) are state-of-the-art models for many computer vision tasks, such as semantic segmentation. A CNN typically stacks a large number of convolution operations to aggregate contextual information~\cite{Simonyan2014vgg}. 
Each convolution is associated with a kernel that consists of a fixed number of learnable weights, which is proportional to the kernel size.
For many semantic segmentation tasks, especially in the biomedical domain, successful CNN architectures integrate both short-range and long-range information~\cite{parsnet,Samy2018nunet}. 
Because convolutions are local operations, successive convolutional layers, increased kernel size, and downsampling operations are often used to capture long-range information and increase capacity~\cite{Krizhevsky2012alex,Lecun1989backprop,unet}.

\begin{figure}[t]
\begin{center}
\includegraphics[width=1\linewidth]{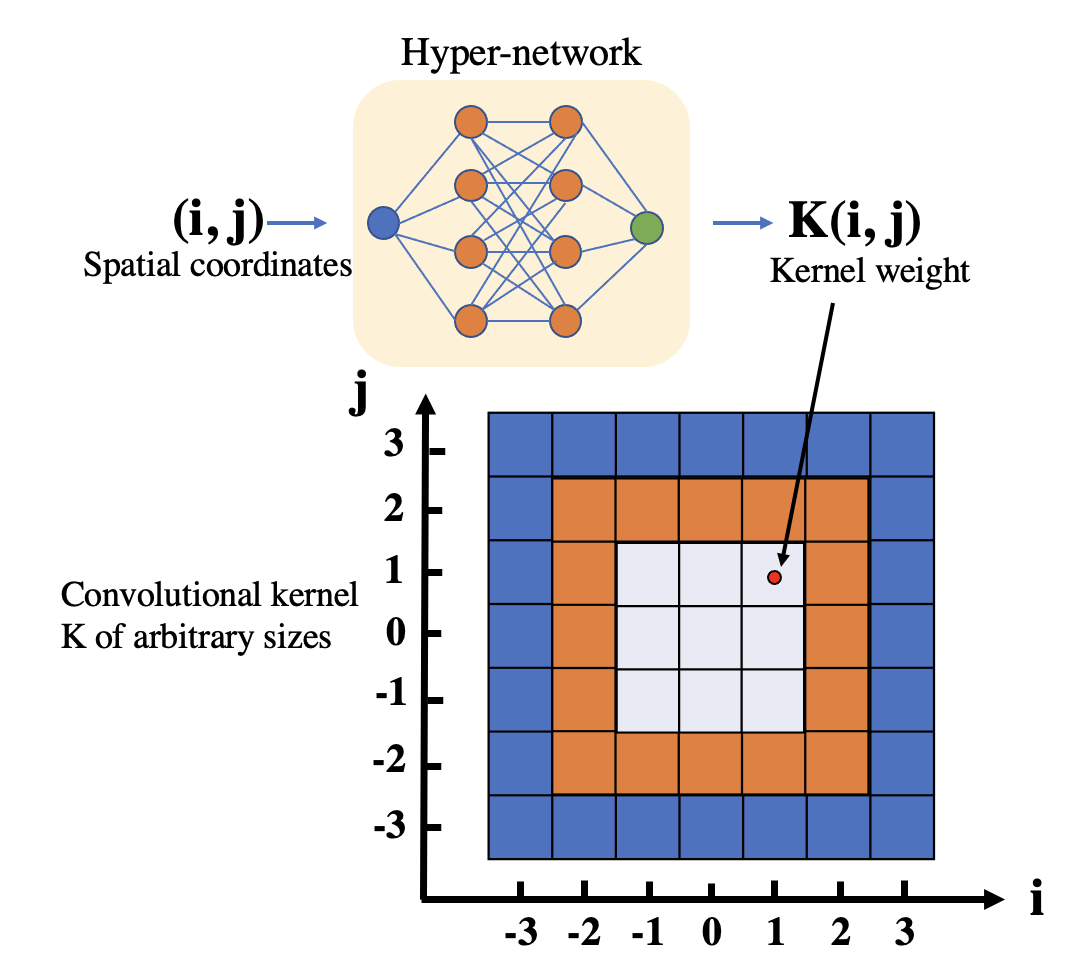}
\end{center}
   \caption{Illustration of the proposed hyper-convolution. A hyper-network takes a coordinate (i,j) and produces the kernel weight at location (i,j) for an arbitrary size convolutional kernel. The only learnable weights are in the hyper-network, independent of the size of the used kernel.
   \vspace{-0.3cm}}
\label{fig:coor}
\end{figure}

A straightforward way to expand the capacity of a CNN is to use kernels with a larger size~\cite{he2019dynamic,large,szegedy2015inception,tan2019mixconv}. However, larger kernels substantially increase the number of learnable parameters, which can lead to overfitting, particularly when training data are limited, as in many biomedical applications.
Alternative representations, such as deformable~\cite{dai2017deformable,thomas2019kpconv} and dilated convolutions ~\cite{dilated3, dilated2,dilated1} can enhance the expressiveness of CNNs. 
However, for tasks such as image segmentation, which requires dense pixel-level classification at the highest resolution, such sparse kernels can be less effective \cite{Huang_2019_CCnet}.

\begin{figure*}[tb]
\begin{center}
%\fbox{\rule{0pt}{2in} \rule{.9\linewidth}{0pt}}
\includegraphics[width=0.9\linewidth]{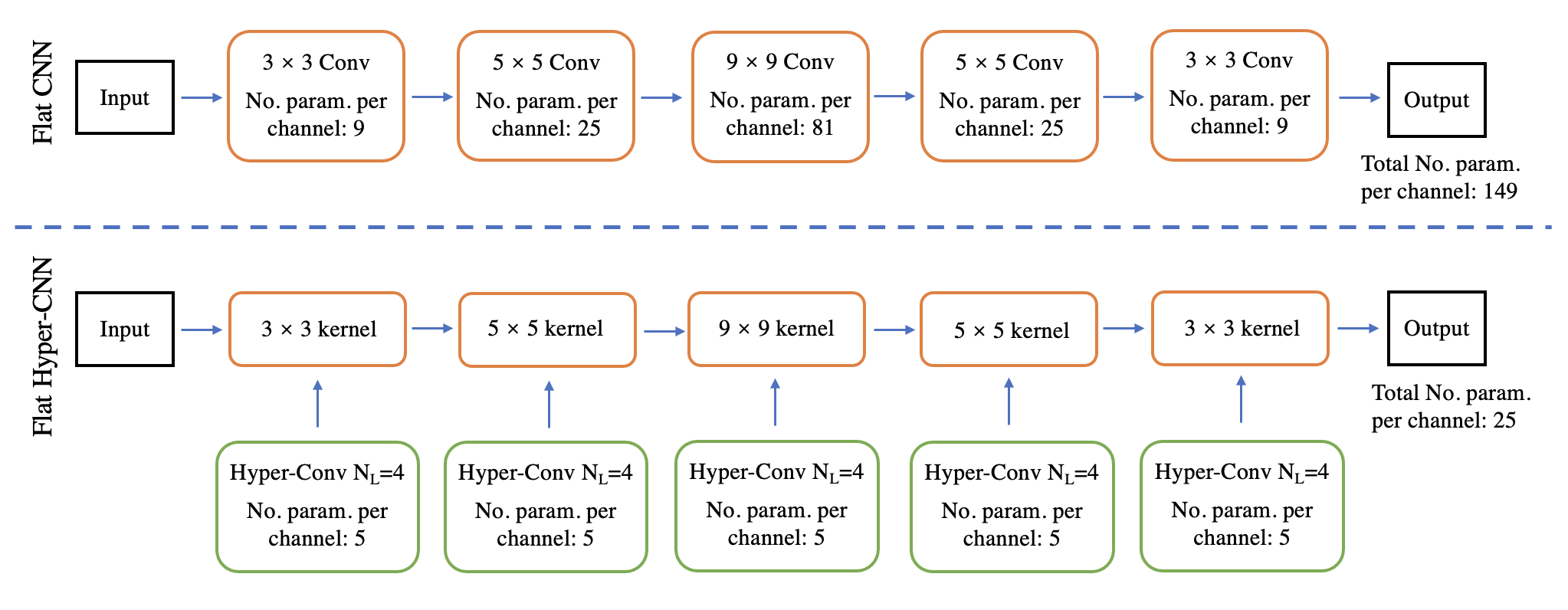}
\vspace{-0.5cm}
\end{center}
   \caption{Overview of the flat CNN (top) and hyper-convolution network (bottom) architectures with same sized kernel for image segmentation. $N_L$ is a hyper-parameter that mostly determines the number of parameters in hyper-network. The flat Hyper-CNN is significantly more parameters efficient.
   \vspace{-0.3cm}}
\label{fig:overview}
\end{figure*}

In this paper, we present a new building block that we call \textit{hyper-convolution}.
A hyper-convolution is an implicit representation of a kernel as a parametrized function of kernel grid coordinates, which decouples the number of learnable parameters from the size of the possible kernel.
We illustrate the hyper-convolution building block in Figure~\ref{fig:coor}.
Similar to a regular convolution with larger kernels, hyper-convolutions can achieve expanded expressiveness but with significantly fewer learnable parameters. 
We perform experiments on two biomedical image segmentation tasks and show improved results with fewer parameters compared to baseline methods. We also observe that the learned hyper-convolutional kernels are naturally spatially regularized, which helps combat overfitting and improves performance. We have analysis of the convolutional kernels in section $4.4$.

The hyper-convolution concept can be used to improve CNN architectures in a variety of ways. They can replace regular convolutions to increase kernel sizes in an existing architecture, given a fixed number of learnable parameters.
Alternatively, hyper-convolutions can reduce the number of learnable parameters in a CNN, without modifying the kernel sizes.
More broadly, the flexibility that comes from decoupling the number of parameters from kernel size enables the design of alternative architectures that might not have been practical with regular convolutions.

A motivating example is illustrated in Figure~\ref{fig:overview}, where a flat CNN implements different kernel sizes without the need for down/up-sampling layers. 
This architecture is competitive for a segmentation problem, yet the substantial number of parameters tends to lead to significant overfitting and optimization challenges. This is one of the main reasons that flat architectures are not used in segmentation problems.
Replacing the convolutions with hyper-convolutions leads to a significantly lower number of parameters, making the architecture less prone to such complications.

Our contribution is two fold: we propose a novel hyper-network based building block for image segmentation that enables flexible kernel designs. The proposed strategy decouples the size and capacity of a convolutional kernel from the total number of parameters. 
Additionally, as we demonstrate empirically, the learned convolutional kernels are naturally regularized and tend to be spatially smooth because the hyper-network encodes kernels as a function of grid coordinates. This property is also desirable in other areas, such as in computer vision tasks where it can afford robustness against adversarial attacks \cite{smooth,wang2020high}. %We believe the hyper-convolution technique can be used as a new regularization for different tasks besides segmentation.

\section{Related Works}
\subsection{Alternative Convolutional Kernels}

There is substantial literature on alternative designs for convolutional building blocks. Large kernels have been explored for image segmentation tasks \cite{large}. Symmetric and separable filters are often employed to reduce the computation cost and the number of parameters.
Dilated convolutions have been widely used to increase the receptive field without increasing the number of learnable parameters~\cite{yu2015dilation}. 
Atrous spatial pyramid pooling was developed to aggregate long-range dependencies in images at multiple scales and yielded excellent performance on several image segmentation datasets \cite{deeplab,dilated3,dilated2}. 
Dilated convolutions have been used in multiple biomedical image segmentation tasks to achieve state-of-the-art results \cite{unetdilated}.
The deformable convolution is another popular technique to increase model capacity by learning sampling locations of kernels and adopting geometric variations in objects \cite{dai2017deformable,zhu2019deformable}.

These powerful techniques maintain a  strong coupling between the number of learnable parameters and the number of neighborhood pixels used in the convolutions. This, in turn, can limit expressiveness for a wide variety of tasks that require dense pixel-level predictions and exhibit long-range dependencies such as biomedical image segmentation. In contrast, our hyper-convolutions break this link, thus enabling the flexibility to design better networks. The hyper-convolution can represent large yet dense kernels that aggregate from all neighborhood pixels, in contrast to the \textit{sparse} kernels adopted by deformable and dilated convolutions with even less parameters.

\subsection{Non-local Network}
Recently, self-attention and non-local networks have gained popularity due to their ability to aggregate long-range information by computing interactions between every pixel pair in a feature map. 
A non-local block gathers contextual information from all other positions in an image by utilizing a self-attention mechanism \cite{attention,nonlocal}. 
Other works built on the non-local architecture and attempted to reduce computational complexity by constructing a more efficient attention map \cite{gnonlocal,asymmetric}. 
Non-local blocks have also been used with a UNet architecture for biomedical image segmentation tasks \cite{nonlocalunet}.
Since non-local operations require substantial computational resources, they are usually only implemented on relatively low-resolution feature maps. In contrast, the proposed hyper-convolution technique is feasible at any resolution.

\subsection{Hyper-Networks}
Hyper-Networks are powerful tools that can improve neural networks' parameter-efficiency without significantly sacrificing expressiveness. The core idea is to use a neural network to generate weights for another network that is responsible for the main task.
For example, Ha et el. \cite{hypernet} used learnable layer embeddings as the input to the hyper-network. For a deep convolutional neural network, this strategy can greatly reduces the number of parameters while maintaining an acceptable performance for a classification task \cite{hypernet}. 
The HyperSeg \cite{nirkin2021hyperseg} architecture encoded the input image and used the encoded features to generate the weights of a decoder that solved a segmentation task. Hyper-networks have also been used to train networks agnostic to the degree of regularization \cite{hyperpara,recon}.

\subsection{Neural Network Implicit Representation}
Neural networks have also been used to create an \textit{implicit} representation of different types of signals, such as natural images. These (usually small) neural networks take in a pixel coordinate and encode an RGB-values image~\cite{represent,Sitzmann2020siren}.
%Similar representations are also used to learn representations of 3D shape~\cite{sal,shape}.
In addition to natural images, people have also used neural networks to learn representations of 3D shape ~\cite{sal,shape}.
More similar to our work, implicit representations have also been used as kernel functions for irregularly structured point cloud data where gridded data are not possible ~\cite{wang2018cloud}.

\section{Proposed Method}
Our core idea is an implicit representation for convolutional kernels.
For a standard convolutional filter, the trainable weights are independent and explicitly learned. Instead, we propose to obtain the value of the kernel given kernel grid coordinates using a parametrized function.
Unlike the standard convolution operation, the size of the convolutional kernel is a design choice that does not affect the number of learnable parameters.

Specifically, a hyper-convolution is a function $\Phi_\theta(\cdot)$ with learnable parameters~$\theta$, that maps kernel grid coordinates to a filter weight $K$. For example, for a 2D condition, 
\begin{equation}
K_{ij} = \Phi_\theta(i,j)
\end{equation}
where $(i,j) \in \mathbb{R}^2$ and $K_{ij}$ indicates the filter weight at filter location $ij$. In our implementation, the center pixel of the convolution kernel has coordinates $(0,0)$.

\subsection{Hyper-network}
We use a neural network to map each 2D input kernel coordinate to the kernel value. The convolutional kernel weights are thus generated by a neural network (hyper-network) instead of independently learned.

For each convolution layer in a segmentation CNN\footnote{Except for the final $1\times1$ convolution layer.}, we implement a corresponding Hyper-CNN $\Phi_\theta(\cdot)$ as a CNN made up of $1\times 1$ convolutional layers, with leaky ReLU nonlinearities with slope of 0.1 \cite{maas2013rectifier}.

Depending on the capacity of the network $\Phi_\theta$, the Hyper-convolutional kernel can be restricted or expressive as a regular convolution kernel. 
In our experiments, we use Hyper-convolution with four hidden layers and the first three layers have a fixed number of nodes. We experiment with several variants of $N_L$, the number of nodes in the final layer.

\subsection{Kernel size and parameter efficiency}
In a standard convolution layer, where the 2D kernel size is $h\times w$ (e.g., $3 \times 3$) and the numbers of input and output channels are $N_{in}$ and $N_{out}$, the total number of parameters is $(h \times w) \times N_{in} \times N_{out}$, excluding the bias terms.  

In the hyper-convolution, a hyper-network with $L$ layers has $(N_L+1) N_{in} N_{out}  +  \sum_{j=0}^{L-1} (N_j+1)N_{j+1}$ parameters, where $N_j$ is the number of nodes in the $j$'th layer. 
Additionally, we have $ N_{in} \times N_{out}$ independent bias terms.
The number of learnable parameters of the hyper-convolution block is independent of the kernel size $h\times w$, and depends on the number of input and output channels, as well as the hyper-parameter $N_L$. 

In practice, $N_{in}$ and $N_{out}$ are most often chosen to be $8$ or larger. Furthermore, in our hyper-network design, we can choose the number of nodes before the penultimate layer $N_L$ to be small (e.g. 8). 
Under these conditions, the number of parameters in the proposed hyper-convolution network is dominated by the final layer and is approximately $(N_L+1) \times N_{in} \times N_{out}$.
If $N_L < h\times w$, the hyper-convolution will have fewer parameters than a standard convolution kernel. 
In this way, for a fixed number of parameters, the proposed representation can implement dense kernels with larger receptive fields, capturing high-resolution contextual information. 

Hyper-convolutions can thus afford expanded receptive field 
%with more capacity 
without increasing the number of training parameters. With a small overhead due to additional operations generating kernel weights, the majority of the memory and computational burden is due to the main network. 
%For example, in our experiments, inference time for $5 \times 5$ UNet and %$5 \times 5$ hyper-UNet are $14$~ms and $16$~ms per slice, respectively.

\subsection{Implementation Details}

The input is a 2-channel pixel coordinate grid of the size equal to the desired kernel size. For instance, for a $3\times 3$ kernel, the input is
%
%-------------------------------------------------------------------------
$$\begin{bmatrix}
$-1$ & $-1$ & $-1$\\
$0$ & $0$ & $0$ \\
$+1$ & $+1$ & $+1$
\end{bmatrix} \begin{bmatrix}
$-1$ & $0$ & $+1$\\
$-1$ & $0$ & $+1$ \\
$-1$ & $0$ & $+1$
\end{bmatrix}.
$$	

In training all methods, we perform data augmentation, including vertical and horizontal flipping, random rotation up to $30$ degrees, and scaling between $0.9$ and $1.1$.  We train all the models using Adam optimizer \cite{adam} with a learning rate of 0.0001 and a mini-batch size of 8 (Liver lesion) or 16 (MS Lesion).  We use dropout regularization with 0.5 probability and batch normalization in all the experiments. 
We use soft Dice loss \cite{milletari2016vnet} for training and report Dice score results for the epoch with the best validation loss. 
The Dice score quantifies the overlap between the automatic and manual segmentations and is widely used in the literature.

%-------------------------------------------------------------------------
\begin{figure}
\begin{center}
%\fbox{\rule{0pt}{2in} \rule{.9\linewidth}{0pt}}
\includegraphics[width=0.9\linewidth]{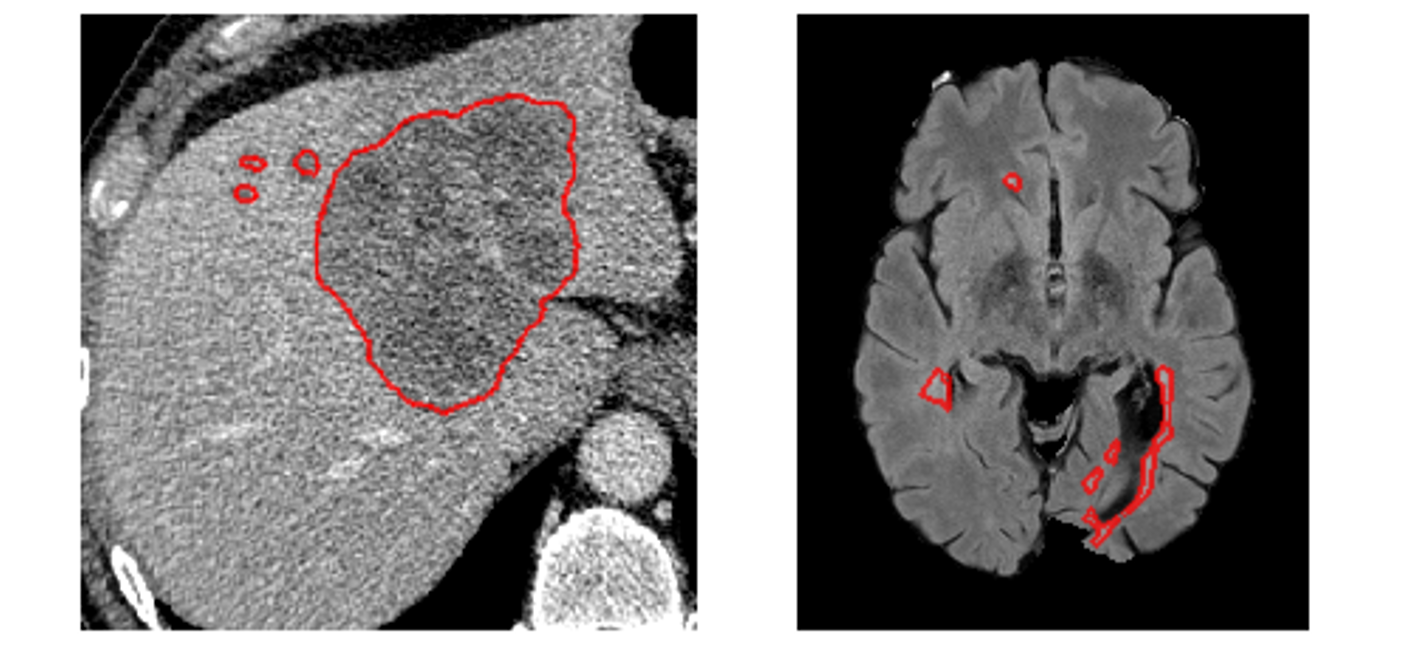}
\vspace{-0.5cm} % just getting us some space
\end{center}
   \caption{Example images and segmentation from the two datasets used in our experiments. Left: Liver lesion. Right: MS-lesion. }
\label{fig:exp}
\end{figure}
%-------------------------------------------------------------------------

\section{Experiments}
The hyper-convolution is a general module that can replace regular convolutions for a wide range of computer vision applications. In this work, we demonstrate it in the context of a segmentation task.
We conduct experiments on two biomedical image segmentation tasks: liver lesion segmentation \cite{lits} and MS lesion segmentation \cite{mslesion} (Figure~\ref{fig:exp}). 
We choose these to demonstrate the hyper-convolution operation since many biomedical image segmentation tasks benefit from larger receptive field while being prone to overfitting.
We then analyze perspectives of the hyper-convolution kernels compared to standard kernels. 

%-------------------------------------------------------------------------
\subsection{Baselines} 
We explore two CNN architectures.
The UNet, a popular architectures used for biomedical segmentation, adopts an multi-scale feature learning, aggregating contextual information using max-pooling operations~\cite{unet}. 
Our 2D UNet backbone has three max-pool layers, two convolution layers per scale, and ReLU nonlinearities. We use a regular $1\times 1$ convolution in the final output layer.
The number of channels is doubled after each max-pool layer. 
Below we indicate the number of channels of the first layer. 
We experiment with varying the kernel size ($3\times 3$ and $5\times 5$), using dilated convolutions~\cite{unetdilated}, and modifying the number of channels, as described in the results sections and supplementary material.
We implement a non-local UNet \cite{nonlocalunet}, which integrates a non-local self-attention block into the bottleneck of the UNet architecture. We also include our implementation of HyperSeg\cite{nirkin2021hyperseg}, a recent segmentation method employing hyper-networks.

The flat CNN backbone \cite{yu2015dilation} consists of a series of convolutional layers with different kernel sizes to expand the receptive field (Figure \ref{fig:overview}). In contrast with the UNet, this architecture does not have any down/up-sampling layers, and convolution operations are executed at the original resolution.
Each network block consists of several convolutions, batch-normalization, and activation.
The kernel sizes in consecutive layers first increase and then decrease, mirroring the UNet contracting and expanding architecture. 
As another baseline, we implement a 2D flat CNN with dilated convolutions that gradually expand the receptive field consisting of sequential residual convolutional kernels of size three, with dilations of 1, 2, 4, 8, 4, 2, 1. 
The numbers of channels for each of these layers are 16, 32, 64, 128, 64, 32, 16, respectively.
%-------------------------------------------------------------------------

\subsection{Liver lesion segmentation}

\textbf{Data:}
We use the LiTS dataset \cite{lits} for liver lesion segmentation, which includes 131 liver CT volumes with ground truth manual segmentation. The number of slices (of size $512\times 512$) in each volume varies between 74 and 987, totaling 58638 2D slices. We resize each slice to $256\times 256$ and truncate the intensity range to $[-100,250]$ before mapping it to $[0,1]$. We randomly split the data and use 80 cases for training, 20 for validation, and 31 for held-out testing.

%-------------------------------------------------------------------------
\begin{figure}[t]
\begin{center}
\includegraphics[width=1\linewidth]{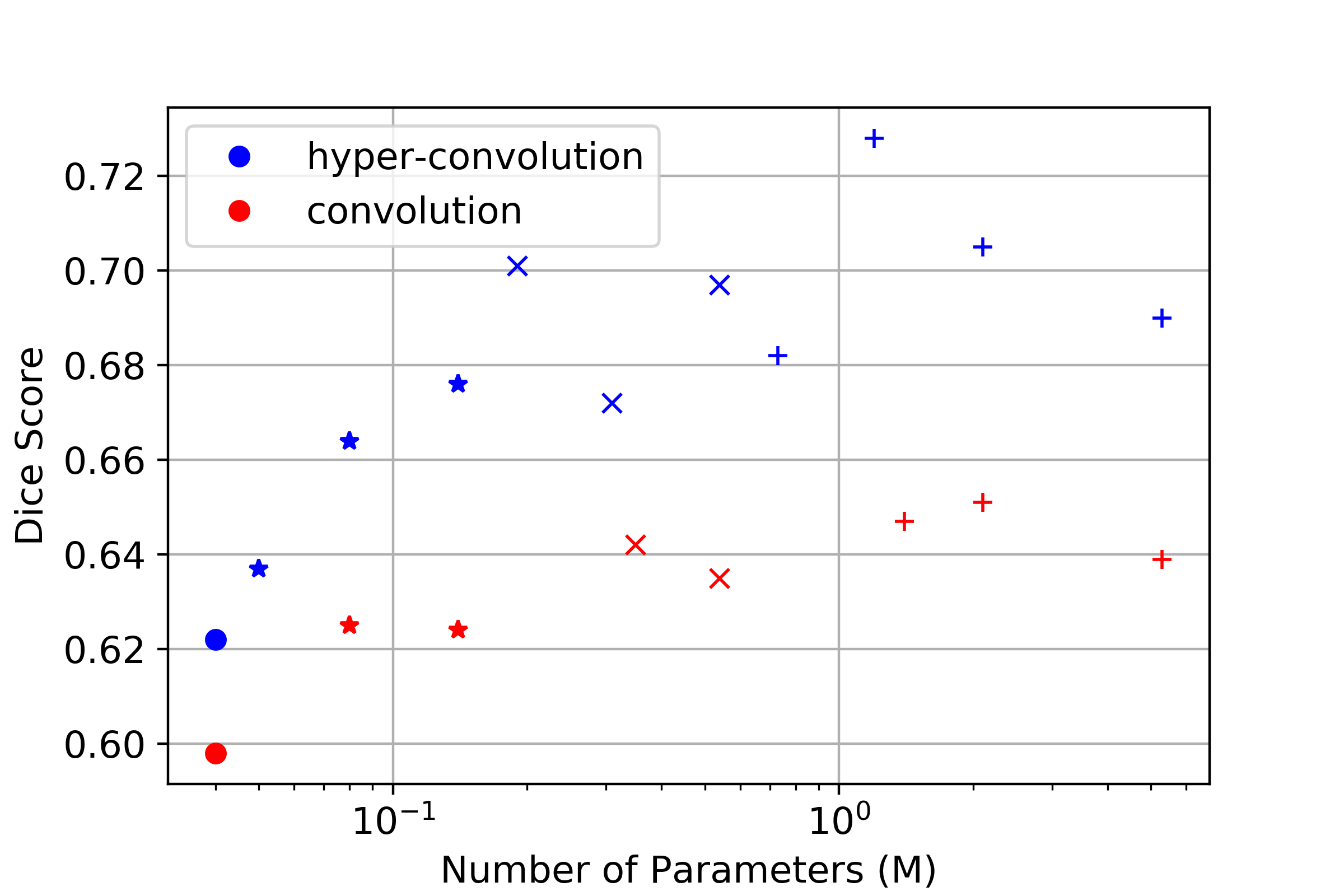}
\end{center}
   \caption{Dice scores for held-out test data segmented
   with $5\times 5$ Hyper-UNet  (blue) and standard UNet (red)  with different numbers of parameters in millions. $\circ$,$\star$,$\times$, and + indicates 4,8,16 and 32 initial channels.}
\label{fig:liver}
\end{figure}
%------------------------------------------------------------------------

\begin{table*}
\begin{center}
\begin{tabular}{|l|c|c|c|c|}
\hline
Method &  Train Dice & Test Dice & Receptive Field & Params (M)\\
\hline\hline
UNet $3\times 3$ \cite{unet} &  0.931 & 0.651 & 68 pixels & 2.1\\
UNet $5\times 5$ &  0.942 & 0.639 & 128 pixels & 5.3\\
Dilated UNet $3\times 3$ \cite{unetdilated} & 0.930 & 0.612 & 128 pixels & 2.1\\
HyperSeg \cite{nirkin2021hyperseg} & 0.902 & 0.683 & All pixels & 1.2 \\
Non-local UNet $3\times 3$ \cite{nonlocalunet} & 0.919 & 0.690 & All pixels & 2.3\\
Hyper-UNet $5\times 5$ (ours) & 0.886 & $0.728$ & 128 pixels & 1.2\\
Non-local Hyper-UNet $5\times 5$ (ours) & 0.893 & $\mathbf{0.733}$ & All pixels & 1.4\\
\hline
Flat Dilated CNN \cite{yu2015dilation} & 0.892 & 0.607 & 89 pixels & 0.45\\
Flat Hyper-CNN (ours) & 0.824 & 0.647 & 89 pixels & 0.45\\
\hline
\end{tabular}
\end{center}
\caption{Train and test performance of different models in the liver lesion segmentation task. Best test Dice score is \textbf{bold-faced}.}\label{tab1}
\vspace{-0.5cm}
\end{table*}
%------------------------------------------------------------------------

\textbf{Results:}
We run extensive experiments with the UNet backbone and the Hyper-UNet versions where all convolutions are replaced with hyper-convolutions of a $5\times5$ kernel size.
For both methods, we vary the number of channels used in the segmentation network, which changes the total number of learnable parameters. 
For the Hyper-UNet, we also vary the number of units $N_L$ in the last layer of the hyper-network. Figure~\ref{fig:liver} shows test Dice scores as the number of parameters are varied for UNet and Hyper-UNet. Hyper-convolutions yield a consistent and significant boost in performance across a wide range of total number of learnable parameters. Dramatically increasing the number of parameters sometimes yields a drop in test performance, likely due to overfitting.  

Table~\ref{tab1} lists training and test results from the epoch with best validation loss, in addition to the receptive field size and number of learnable parameters, for a collection of baseline models and their hyper-convolution counterparts.
As we increase the kernel size from 3 to 5 in the UNet baseline, the train Dice improves, indicating a model with better expressiveness. However, with this change, the total number of learnable parameters more than double, and the UNet with the larger kernel exhibits more overfitting, as evidenced by the increased difference between the test and train Dice scores. 
The dilated UNet baseline with a kernel size of 3 achieves the same receptive field as a $5\times 5$ UNet. 
Despite having the same number of learnable parameters as a $3\times 3$ UNet, the dilated UNet has worse test performance, indicating that simply increasing the receptive field does not necessarily improve model performance. 
By utilizing a self-attention mechanism, the non-local UNet baseline shows a robust improvement in test Dice score, without a significant increase in the total number of parameters. 
The performance boost is likely due to its capability of aggregating information from all pixels and thus utilizing global information to make predictions. 

For methods with hyper-networks, we replace regular convolution in the baselines with hyper-convolution. The Hyper-UNet and non-local Hyper-UNet modify the UNet and non-local UNet backbones by replacing all convolutions with $5 \times 5$ hyper-convolutions that have $N_L=4$. These hyper-convolutions in Hyper-UNet provide an increased receptive field while only having half of the total number of learnable parameters as the $3\times3$ UNet baseline. 
With a more restricted kernel as an effective regularization, the proposed Hyper-UNet and non-local Hyper-UNet show less overfitting and achieve the best test performance compared to all other methods. 

The flat CNN baseline has 0.45M of learnable parameters, which is much less than the $3\times3$ UNet, but has a receptive field that is larger. 
The test performance, however, is the worst among all baselines, due to substantial overfitting.
In contrast, the flat Hyper-CNN, which implements dense kernels with the same receptive field size as the flat  baseline, achieves a test Dice score that is comparable to the $3\times3$ UNet - with a significant reduction in the gap between test and train scores and 1/5'th of the number of parameters - the same as the flat baseline. 
This result suggests that it is not just the reduced number of parameters in the hyper-convolution that yields better test performance. 
In fact, as we show below, we believe that the spatial regularization achieved by hyper-convolution kernels can, in part, explain the performance boost.

%------------------------------------------------------------------------、

\begin{table}
\begin{center}
\begin{tabular}{|c|c|c|c|}
\hline
Size, $N_L$  & Test Dice & Recep. Field & Params (M)\\
\hline\hline
$3\times 3$, $2$  & 0.604 & 68 pixels & 0.73\\
$3\times 3$, $4$  & 0.627 & 68 pixels & 1.2\\
$3\times 3$, $8$  & 0.648 & 68 pixels & 2.2\\
\hline
$5\times 5$, $2$  & 0.692 & 128 pixels & 0.73\\
$5\times 5$, $4$  & 0.728 & 128 pixels & 1.2\\
$5\times 5$, $8$  & 0.705 & 128 pixels & 2.2\\
\hline
$7\times 7$, $2$  & 0.683 & 188 pixels & 0.73\\
$7\times 7$, $4$  & 0.704 & 188 pixels & 1.2\\
$7\times 7$, $8$  & 0.717 & 188 pixels & 2.2\\

\hline
\end{tabular}
\end{center}
\caption{Performance of Hyper-UNet on Liver Lesion data with different kernel sizes and hyper-network capacity.}\label{tab2}
\vspace{-0.5cm}
\end{table}
%------------------------------------------------------------------------

\begin{figure}[t]
\begin{center}
\includegraphics[width=1\linewidth]{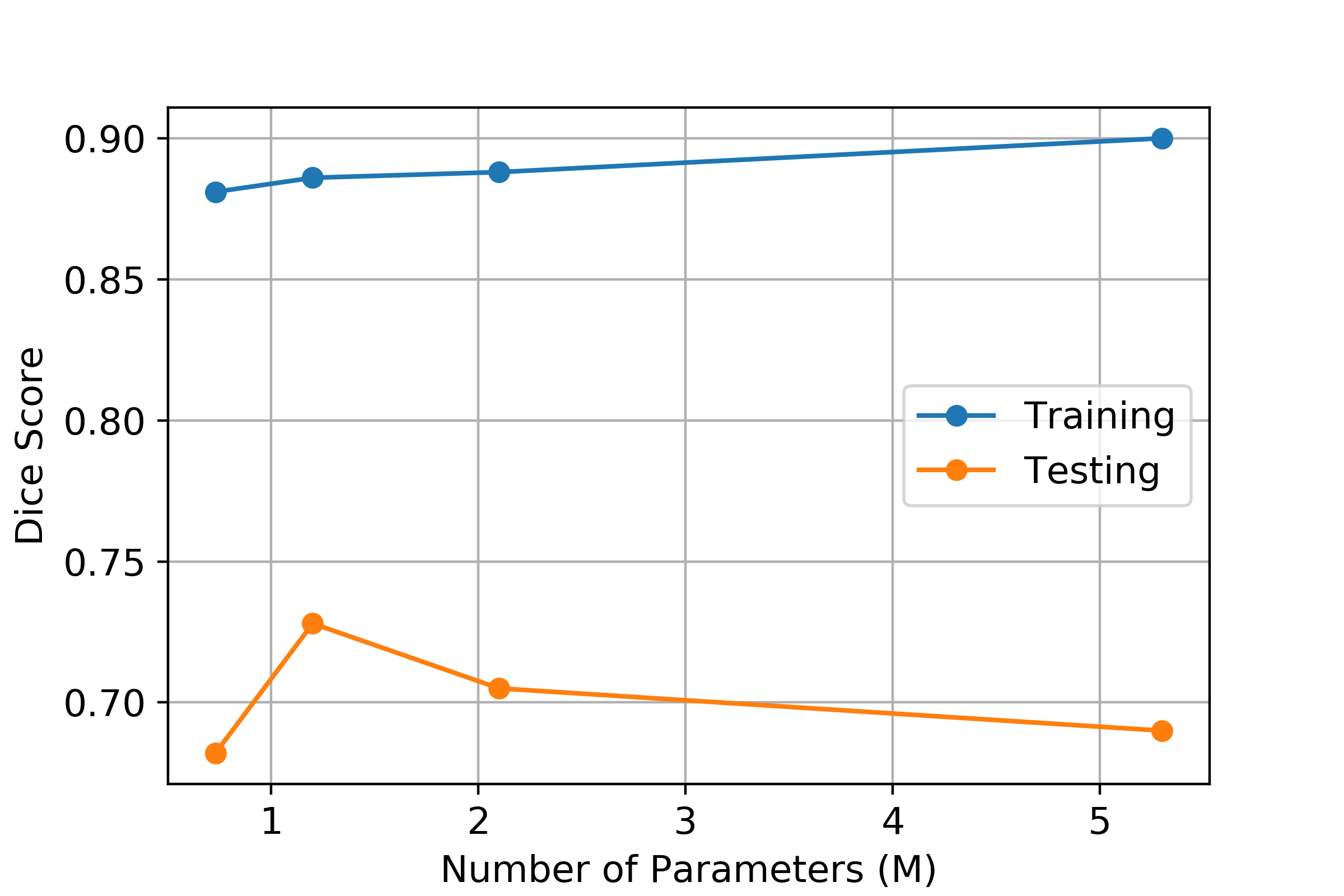}
\vspace{-0.5cm}
\end{center}
   \caption{Train (blue) and test (orange) Dice scores for $5\times 5$ Hyper-UNet with different numbers of parameters in millions, corresponding to $N_L = 2, 4, 8,  24$.
   \vspace{-0.5cm}}
\label{fig:traintest}
\end{figure}
%------------------------------------------------------------------------

\textbf{Hyperparameters:} 
Table~\ref{tab2} shows results for the 32-channel Hyper-UNet with variable kernel sizes and $N_L$ values. We observe that the $3\times 3$ hyper-convolution kernel performs worse than the standard $3\times 3$ convolution, possibly because of its restricted capacity. The gap shrinks as we increase $N_L$. 
Larger kernel sizes achieve better test Dice scores, however increasing hyper-network capacity (i.e., $N_L$) does not always improve performance (seen with $5\times5$ kernels), likely due to overfitting. 

Figure~\ref{fig:traintest} shows the train and test Dice scores for $5\times 5$ hyper-convolutions with different $N_L=2, 4, 8, 24$. 
A hyper-convolution model with $N_L=24$ has approximately the same number of learnable parameters as a regular $3 \times 3$ UNet. We note that the train Dice increases with number of parameters, indicating better model expressiveness. However, the test Dice peaks at $N_L=4$, demonstrating that regularization via restricted hyper-convolution capacity can improve generalization. 

%------------------------------------------------------------------------

\subsection{Multiple Sclerosis Lesion Segmentation}

\textbf{Data:}
Next, we consider a Multiple Sclerosis (MS) lesion segmentation task.
We use a public dataset~\cite{mslesion}, which contains brain MRI scans from 19 subjects, each with 4-6 scans from different time points. Among the 19 subjects, manual annotations are provided for 5 subjects (21 images), and the remaining 14 subjects (61 images) are used for independent testing. 
For the 14 test subjects, we submit segmentations via an online portal for evaluation, which reports back test Dice scores.
The dataset has two independent expert annotations, which have a Dice overlap score of 0.732, highlighting how challenging the task is. We use the intersection of the two manual labels as the gold standard labels during training and validation. 
Each image contains four different Magnetic Resonance Imaging (MRI) contrasts: FLAIR, PD-weighted, T2-weighted, and T1-weighted. The original images have size $182 \times 256 \times 182$. 
In total there are 3822 2D images for training. 
We crop the center of each 3D image to $144 \times 176 \times 144$ and apply z-score normalization for subsequent training. 

Since we only have 5 subjects with gold standard segmentations, we run 5-fold experiments where each fold has 4 train subjects and 1 validation subject. All reported Dice scores (train and test) are averaged across these 5 folds. As the baseline UNet model, we adopt the multi-branch variant \cite{aslani2019multi} (MB-UNet) which utilizes the 4 modalities and all orthogonal planes of the 3D volumes to achieve competitive results for this task.    
%------------------------------------------------------------------------

\begin{figure}[tb]
\begin{center}
\includegraphics[width=1\linewidth]{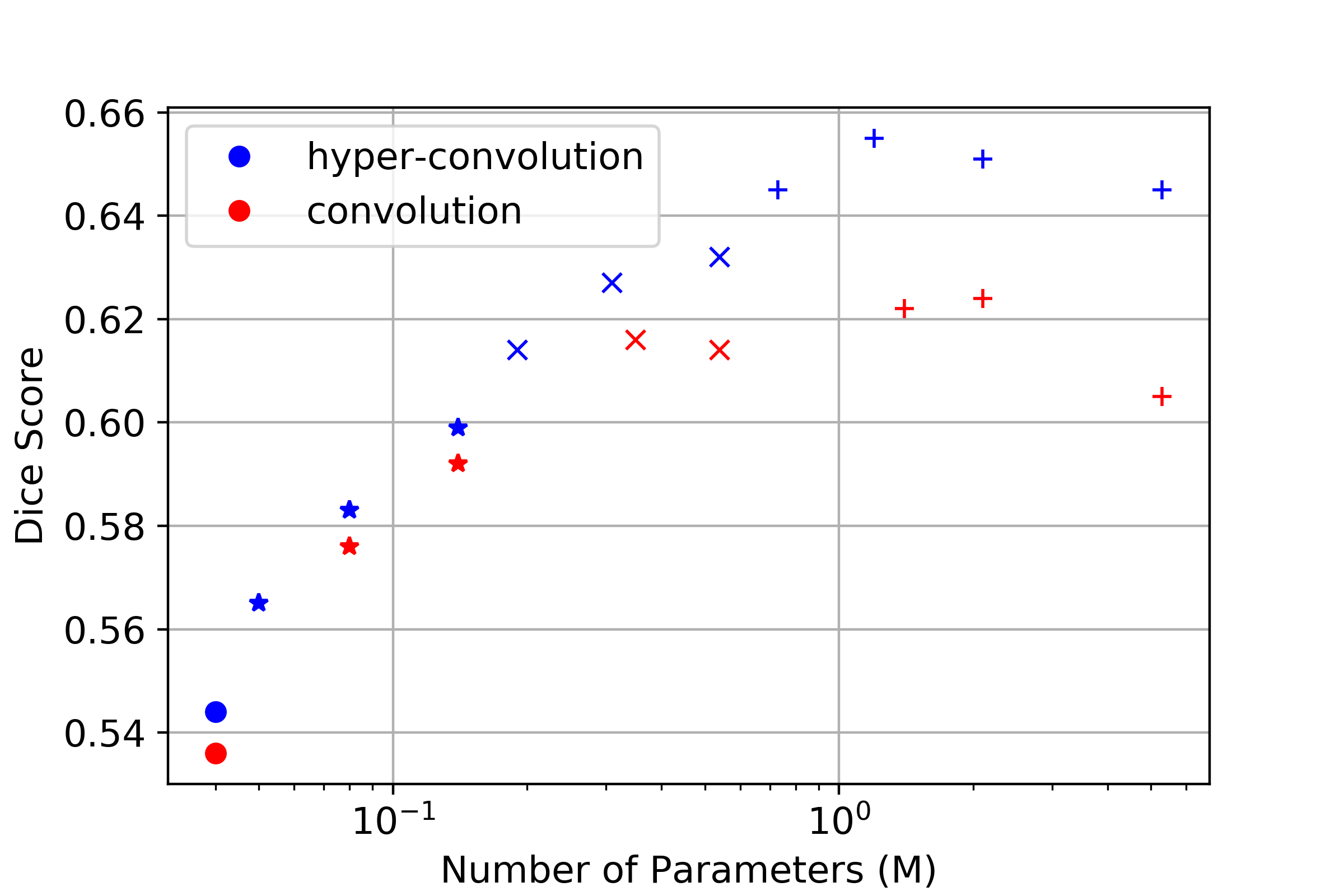}
\end{center}
   \caption{Test Dice scores for $5\times 5$ Hyper-UNet (blue) and UNet baseline (red) with different numbers of parameters. $\circ$,$\star$,$\times$, and + indicates 4,8,16 and 32 initial channels.}
\label{brain}
\end{figure}
%------------------------------------------------------------------------

\begin{table*}
\begin{center}
\begin{tabular}{|l|c|c|c|c|}
\hline
Method &  Train Dice & Test Dice & Receptive Field & Params (M)\\
\hline\hline
MB-UNet $3\times 3$\cite{aslani2019multi} &  0.887 & 0.624 & 68 pixels & 2.1\\
MB-UNet $5\times 5$ &  0.893 & 0.605 & 128 pixels & 5.3\\
Dilated MB-UNet $3\times 3$ & 0.881 & 0.625 & 128 pixels & 2.1\\
HyperSeg \cite{nirkin2021hyperseg} & 0.864 & 0.64 & All pixels & 1.2 \\
Non-local MB-UNet $3\times 3$ \cite{nonlocalunet} & 0.905 & 0.637 & All pixels & 2.3\\
Hyper-MB-UNet $5\times 5$ (ours) & 0.82 & $\mathbf{0.655}$ & 128 pixels & 1.2\\
Non-local Hyper-MB-UNet $5\times 5$ (ours) & 0.834 & $0.652$ & All pixels & 1.4\\
\hline
Flat Dilated CNN\cite{yu2015dilation} & 0.854 & 0.616 & 89 pixels & 0.45\\
Flat Hyper-CNN (ours) & 0.805 & 0.649 & 89 pixels & 0.45\\
\hline
\end{tabular}
\end{center}
\caption{Train and Test Performance of different models in MS lesion segmentation task. Best test Dice score is \textbf{bold-faced}.}\label{tab3}
\end{table*}
%------------------------------------------------------------------------

\textbf{Results:}
Figure~\ref{brain} shows test Dice scores for the MB-UNet models with different numbers of parameters obtained by varying the number of channels and kernel size for standard convolution. 
The hyper-convolutions implement $5\times5 $ kernels.
As in the liver lesion segmentation task, we observe that hyper-convolutions consistently boost the test Dice score over a wide range of parameterizations.  

Table~\ref{tab3} lists train and test results from the epoch with best validation loss, in addition to the receptive field size and number of learnable parameters, for baseline models and their hyper-convolution counterparts.
Similar to above, we observe that hyper-convolutions boost test performance and shrink the gap between train and test loss. 

Contrary to what we observed in liver lesion segmentation, the flat Hyper-CNN yields better results than the BM-UNet baseline and the non-local UNet, which have larger receptive fields. We believe this difference can be attributed to the fact that MS lesions are relatively small compared to liver lesions, and thus their segmentation does not require a large receptive field (Figure~\ref{fig:exp}).  

\textbf{Hyperparameters:} 
Table~\ref{tab4} shows results for the 32-channel Hyper-MB-UNet with variable kernel sizes and $N_L$ values. We observe that the $5\times 5$ hyper-convolution kernel yields the best results.
We also note that, as before, increasing the capacity of the hyper-network does not always yield better test performance, presumably due to overfitting. 
This underscores the importance of the regularization achieved by using restricted hyper-convolutions.

\begin{table}[b]
\begin{center}
\begin{tabular}{|c|c|c|c|}
\hline
Size, $N_L$ & Test Dice & Recep. Field & Params (M)\\
\hline\hline
$3\times 3$, $2$ & 0.622 & 68 pixels & 0.73\\
$3\times 3$, $4$ & 0.625 & 68 pixels & 1.2\\
$3\times 3$, $8$ & 0.617 & 68 pixels & 2.2\\
\hline
$5\times 5$, $2$ & 0.648 & 128 pixels & 0.73\\
$5\times 5$, $4$ & 0.655 & 128 pixels & 1.2\\
$5\times 5$, $8$ & 0.651 & 128 pixels & 2.2\\
\hline
$7\times 7$, $2$ & 0.634 & 188 pixels & 0.73\\
$7\times 7$, $4$ & 0.644 & 188 pixels & 1.2\\
$7\times 7$, $8$ & 0.646 & 188 pixels & 2.2\\

\hline
\end{tabular}
\end{center}
\caption{Performance of Hyper-MB-UNet on MS Lesion data with different kernel sizes and hyper-network capacity.}\label{tab4}
\end{table}

\subsection{Kernel Smoothness}
%------------------------------------------------------------------------

\begin{figure*}[tb]
\begin{center}
\includegraphics[width=1\linewidth]{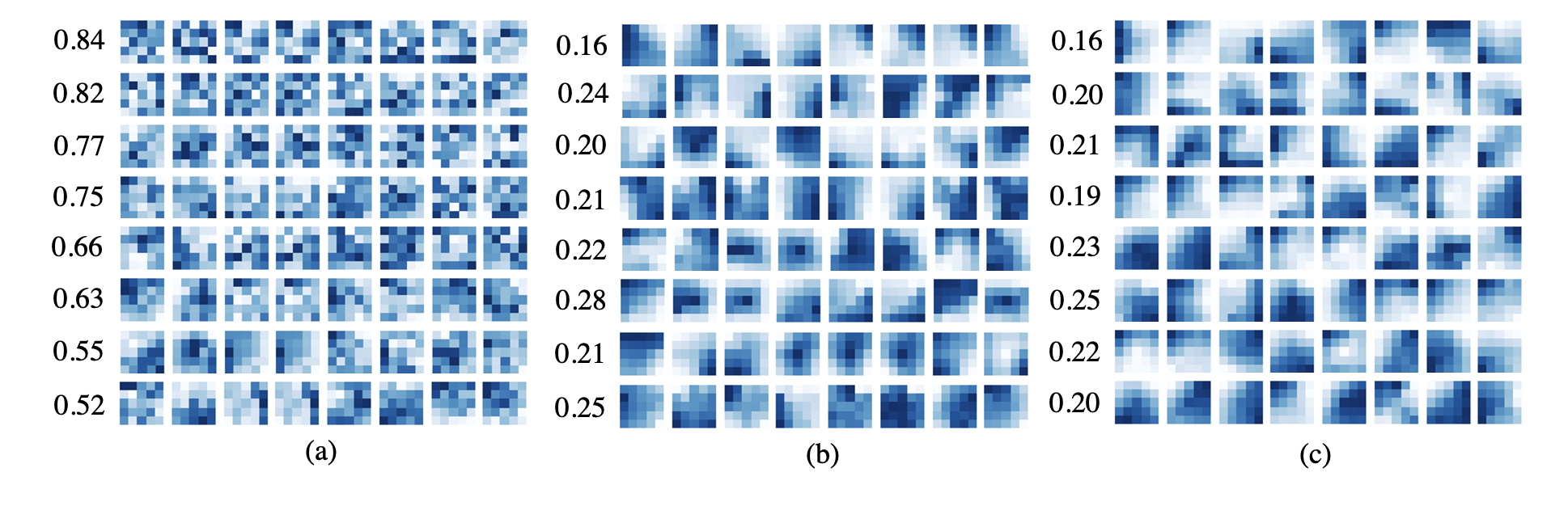}
\vspace{-1.1cm} % just getting us some space
\end{center}
   \caption{Visualization of $5\times 5$ convolutional kernels and their Laplacians in different layers of the networks for (a) UNet baseline. (b) Hyper-UNet ($N_L=8$). (c) Hyper-UNet ($N_L=24$). Each row corresponds to one network layer, with the average Laplacian value listed.}
\label{fig:kernel1}
\end{figure*}
%------------------------------------------------------------------------

\begin{figure*}[tb]
\begin{center}
\includegraphics[width=0.7\linewidth]{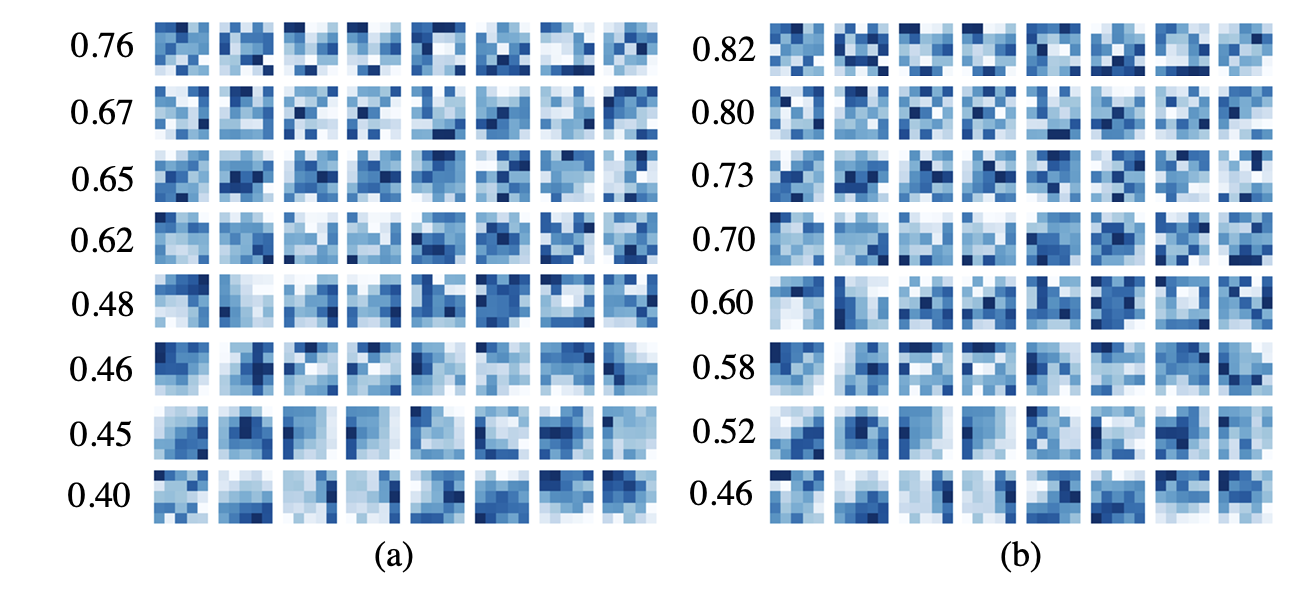}
\vspace{-0.5cm} % just getting us some space
\end{center}
   \caption{Visualizations of reconstructed $5\times 5$ regular UNet kernels from Figure~\ref{fig:kernel1}(a) using a Hyper-CNN with (a) $N_L=8$. (b) $N_L=24$. Each row corresponds to one network layer with the average Laplacian value listed.}
\label{fig:kernel2}
\end{figure*}
%------------------------------------------------------------------------

To gain further insights into hyper-convolutions, we perform an analysis of the kernels of the $5\times5$ UNet and Hyper-UNet models trained for the liver segmentation task. 
Figure~\ref{fig:kernel1} shows the learned $5\times 5$ kernels for standard convolutions and hyper-convolutions. 
In each panel, each row corresponds to one layer in the network, from which we show 8 randomly chosen kernels. 
We observe that the kernels learned in both the low capacity ($N_L=8$) and high capacity ($N_L=24$) Hyper-CNNs are significantly smoother than those learned in the standard UNet, despite the high capacity hyper-convolution being equally expressive as the standard $5\times5$ convolution (as we show below).
The smoothness of a kernel can be quantified by calculating its average 2nd-order spatial derivative (Laplacian). Lower Laplacian values indicate smoother kernels.
Layer-wise average Laplacian values are listed in Figure~\ref{fig:kernel1}, which corroborate our visual assessment. 

To better understand whether the smoothness of the hyper-convolution kernels is due to learning dynamics or limited expressiveness, we experimented with mapping the learned regular convolution kernels directly to hyper-convolution kernels. To achieve this, we train a hyper-network to reconstruct each learned UNet kernel by minimizing L2 loss on kernel weight values. As we can see from Figure~\ref{fig:kernel2}, both the low ($N_L=8$) and high capacity ($N_L=24$) hyper-networks can well approximate the CNN kernels shown in Figure~\ref{fig:kernel1}(a) and achieve similar Laplacian values. 
While the high capacity yields a more accurate representation, the low capacity hyper-network is surprisingly close too, even though
it has less than half of the parameters of the regular $5\times 5$ convolution.

These results show that the smooth kernels learned by the Hyper-CNN (Figure~\ref{fig:kernel1}) are not due to the limited capacity of the hyper-network, but rather learning dynamics.   

The benefits of smooth kernels have been studied in previous research. Feinman and Lake \cite{smooth} proposed to use a smooth kernel regularizer to encourage the kernel weights to be spatially correlated. They demonstrated that smooth kernels show better generalization performance. Recently, Wang \etal \cite{wang2020high} also suggested that models with smooth convolutional kernels, especially in the earlier layers, tend to have better adversarial robustness. They argued that a smooth kernel can ignore the high-frequency component of an image which is usually invisible to the human eye and can be disruptive for predictions. 
%In hyper-convolutions, the spatial smoothness in the learned kernels can be directly imposed by restricting the capacity of the hyper-network, yet we observe that even with high capacity the learned kernels exhibit spatial smoothness. 
We believe that this spatial smoothness explains the smaller gap between the train and test performance we observe with Hyper-CNNs. 

We also find that deeper layer kernels in the Hyper-CNN are less smooth than the first layer, as evidenced by higher Laplacian values. For the convolution that is directly operated on the input image, smooth kernels can eliminate the noise and make the model more robust. 
%In our baseline CNN models, We do not observe such a pattern. Instead, the smoothness of kernels seems to increase with deeper layers. 
We include a more detailed analysis of this in the Supplementary Material.

\section{Conclusion}
In this paper, we presented hyper-convolution, a novel building block that can be used with any convolutional neural network architecture. The hyper-convolution represents kernel weights as an implicit function of grid coordinates, as opposed to regular convolutions that treat each kernel weight independently. Hyper-convolutions decouple the total number of learnable parameters in a kernel from its size, enabling us to use larger filters with greater receptive field without having too many learnable parameters. It can also be used to reduce the total number of parameters without modifying the receptive field but allowing for regularization. We observe that the learned hyper-convolution kernels are smoother than their regular counterparts, which can help combat overfitting and improve generalization and robustness. 
%Compared with techniques such as dilated and deformable convolutions, the hyper-convolution can capture richer contextual information within its field-of-view. 

We conducted experiments on two challenging biomedical tasks: liver lesion segmentation and MS-lesion segmentation. 
%We use both a UNet and a flat CNN as backbones to test the performance of hyper-convolution against other competing methods including dilated convolution and non-local network.
We test the performance of hyper-convolution against other competing methods including dilated convolution and non-local network. 
We demonstrated that hyper-convolutions can boost performance by increasing the receptive field, reducing the number of learnable parameters, and/or regularizing the kernels. 
We believe hyper-convolutions will be an important building block for future neural network architectures, enabling researchers to further explore the trade-offs between capacity and generalization. 

\section*{Acknowledgement}
This work was supported by NIH grants R01LM012719, R01AG053949, the NSF NeuroNex grant 1707312, and the NSF CAREER 1748377 grant.

\newpage
{\small
\bibliographystyle{ieee_fullname}
\bibliography{egbib}
}

\end{document}